\documentclass[10pt,aps,prl,twocolumn,showpacs,superscriptaddress,floatfix,preprintnumbers,amsmath,amssymb,booktab]{revtex4-1}  

\usepackage{graphicx}
\usepackage{dcolumn}
\usepackage{booktabs}
\usepackage{array}
\usepackage{bm}
\usepackage{natbib}
\usepackage{amssymb}   
\usepackage[pagewise]{lineno}
\usepackage[usenames,dvipsnames]{color}

\begin{document}
\preprint{APS/123-QED}

\title{Search for neutrinos from annihilation of captured low-mass dark matter particles in the Sun by Super-Kamiokande}
\newcommand{\AFFicrr}{\affiliation{Kamioka Observatory, Institute for Cosmic Ray Research, University of Tokyo, Kamioka, Gifu 506-1205, Japan}}
\newcommand{\AFFkashiwa}{\affiliation{Research Center for Cosmic Neutrinos, Institute for Cosmic Ray Research, University of Tokyo, Kashiwa, Chiba 277-8582, Japan}}
\newcommand{\AFFipmu}{\affiliation{Kavli Institute for the Physics and
Mathematics of the Universe (WPI), Todai Institutes for Advanced Study,
University of Tokyo, Kashiwa, Chiba 277-8582, Japan }}
\newcommand{\AFFmad}{\affiliation{Department of Theoretical Physics, University Autonoma Madrid, 28049 Madrid, Spain}}
\newcommand{\AFFubc}{\affiliation{Department of Physics and Astronomy, University of British Columbia, Vancouver, BC, V6T1Z4, Canada}}
\newcommand{\AFFbu}{\affiliation{Department of Physics, Boston University, Boston, MA 02215, USA}}
\newcommand{\AFFbnl}{\affiliation{Physics Department, Brookhaven National Laboratory, Upton, NY 11973, USA}}
\newcommand{\AFFuci}{\affiliation{Department of Physics and Astronomy, University of California, Irvine, Irvine, CA 92697-4575, USA }}
\newcommand{\AFFcsu}{\affiliation{Department of Physics, California State University, Dominguez Hills, Carson, CA 90747, USA}}
\newcommand{\AFFcnm}{\affiliation{Department of Physics, Chonnam National University, Kwangju 500-757, Korea}}
\newcommand{\AFFduke}{\affiliation{Department of Physics, Duke University, Durham NC 27708, USA}}
\newcommand{\AFFfukuoka}{\affiliation{Junior College, Fukuoka Institute of Technology, Fukuoka, Fukuoka 811-0295, Japan}}
\newcommand{\AFFgmu}{\affiliation{Department of Physics, George Mason University, Fairfax, VA 22030, USA }}
\newcommand{\AFFgifu}{\affiliation{Department of Physics, Gifu University, Gifu, Gifu 501-1193, Japan}}
\newcommand{\AFFgist}{\affiliation{GIST College, Gwangju Institute of Science and Technology, Gwangju 500-712, Korea}}
\newcommand{\AFFuh}{\affiliation{Department of Physics and Astronomy, University of Hawaii, Honolulu, HI 96822, USA}}
\newcommand{\AFFkanagawa}{\affiliation{Physics Division, Department of Engineering, Kanagawa University, Kanagawa, Yokohama 221-8686, Japan}}
\newcommand{\AFFkek}{\affiliation{High Energy Accelerator Research Organization (KEK), Tsukuba, Ibaraki 305-0801, Japan }}
\newcommand{\AFFkobe}{\affiliation{Department of Physics, Kobe University, Kobe, Hyogo 657-8501, Japan}}
\newcommand{\AFFkyoto}{\affiliation{Department of Physics, Kyoto University, Kyoto, Kyoto 606-8502, Japan}}
\newcommand{\AFFumd}{\affiliation{Department of Physics, University of Maryland, College Park, MD 20742, USA }}
\newcommand{\AFFmit}{\affiliation{Department of Physics, Massachusetts Institute of Technology, Cambridge, MA 02139, USA}}
\newcommand{\AFFmiyagi}{\affiliation{Department of Physics, Miyagi University of Education, Sendai, Miyagi 980-0845, Japan}}
\newcommand{\AFFnagoya}{\affiliation{Solar Terrestrial Environment Laboratory, Nagoya University, Nagoya, Aichi 464-8602, Japan}}
\newcommand{\AFFpol}{\affiliation{National Centre For Nuclear Research, 00-681 Warsaw, Poland}}
\newcommand{\AFFsuny}{\affiliation{Department of Physics and Astronomy, State University of New York at Stony Brook, NY 11794-3800, USA}}
\newcommand{\AFFniigata}{\affiliation{Department of Physics, Niigata University, Niigata, Niigata 950-2181, Japan }}
\newcommand{\AFFokayama}{\affiliation{Department of Physics, Okayama University, Okayama, Okayama 700-8530, Japan }}
\newcommand{\AFFosaka}{\affiliation{Department of Physics, Osaka University, Toyonaka, Osaka 560-0043, Japan}}
\newcommand{\AFFregina}{\affiliation{Department of Physics, University of Regina, 3737 Wascana Parkway, Regina, SK, S4SOA2, Canada}}
\newcommand{\AFFseoul}{\affiliation{Department of Physics, Seoul National University, Seoul 151-742, Korea}}
\newcommand{\AFFshizuokasc}{\affiliation{Department of Informatics in
Social Welfare, Shizuoka University of Welfare, Yaizu, Shizuoka, 425-8611, Japan}}
\newcommand{\AFFskk}{\affiliation{Department of Physics, Sungkyunkwan University, Suwon 440-746, Korea}}
\newcommand{\AFFtohoku}{\affiliation{Research Center for Neutrino Science, Tohoku University, Sendai, Miyagi 980-8578, Japan}}
\newcommand{\AFFtokyo}{\affiliation{The University of Tokyo, Bunkyo, Tokyo 113-0033, Japan }}
\newcommand{\AFFtoronto}{\affiliation{Department of Physics, University of Toronto, 60 St., Toronto, Ontario, M5S1A7, Canada }}
\newcommand{\AFFtriumf}{\affiliation{TRIUMF, 4004 Wesbrook Mall, Vancouver, BC, V6T2A3, Canada }}
\newcommand{\AFFtokai}{\affiliation{Department of Physics, Tokai University, Hiratsuka, Kanagawa 259-1292, Japan}}
\newcommand{\AFFtit}{\affiliation{Department of Physics, Tokyo Institute
for Technology, Meguro, Tokyo 152-8551, Japan }}
\newcommand{\AFFtsinghua}{\affiliation{Department of Engineering Physics, Tsinghua University, Beijing, 100084, China}}
\newcommand{\AFFwarsaw}{\affiliation{Institute of Experimental Physics, Warsaw University, 00-681 Warsaw, Poland }}
\newcommand{\AFFuw}{\affiliation{Department of Physics, University of Washington, Seattle, WA 98195-1560, USA}}

\AFFicrr
\AFFkashiwa
\AFFmad
\AFFbu
\AFFubc
\AFFbnl
\AFFuci
\AFFcsu
\AFFcnm
\AFFduke
\AFFfukuoka
\AFFgifu
\AFFgist
\AFFuh
\AFFkek
\AFFkobe
\AFFkyoto
\AFFmiyagi
\AFFnagoya
\AFFpol
\AFFsuny
\AFFokayama
\AFFosaka
\AFFregina
\AFFseoul
\AFFshizuokasc
\AFFskk
\AFFtokai
\AFFtokyo
\AFFipmu
\AFFtoronto
\AFFtriumf
\AFFtsinghua
\AFFuw

\author{K.~Choi}
\AFFnagoya
\author{K.~Abe}
\AFFicrr
\AFFipmu
\author{Y.~Haga}
\AFFicrr
\author{Y.~Hayato}
\AFFicrr
\AFFipmu
\author{K.~Iyogi}
\AFFicrr 
\author{J.~Kameda}
\author{Y.~Kishimoto}
\author{M.~Miura} 
\author{S.~Moriyama} 
\author{M.~Nakahata}
\AFFicrr
\AFFipmu 
% ...

\collaboration{The Super-Kamiokande Collaboration}
\noaffiliation

\date{\today}

\begin{abstract}
Super-Kamiokande (SK) can search for weakly interacting massive particles~(WIMPs) by detecting neutrinos produced from WIMP annihilations occurring inside the Sun. In this analysis, we include neutrino events with interaction vertices in the detector in addition to upward-going muons produced in the surrounding rock. Compared to the previous result, which used the upward-going muons only, the signal acceptances for light (few-GeV/$c^2$ $\sim$ 200-GeV/$c^2$) WIMPs are significantly increased.
We fit 3903 days of SK data to search for the contribution of neutrinos from WIMP annihilation in the Sun.
We found no significant excess over expected atmospheric-neutrino background and the result is interpreted in terms of upper limits on WIMP-nucleon elastic scattering cross sections under different assumptions about the annihilation channel.
We set the current best limits on the spin-dependent~(SD) WIMP-proton cross section for WIMP masses below 200 GeV/$c^2$ (at 10 GeV/$c^2$, 1.49$\times 10^{-39}$~cm$^2$ for $\chi\chi\rightarrow b \overline{b}$ and 1.31$\times 10^{-40}$~cm$^2$ for $\chi\chi\rightarrow\tau^+\tau^-$ annihilation channels), also ruling out some fraction of WIMP candidates with spin-independent~(SI) coupling in the few-GeV/$c^2$ mass range.
\end{abstract}

\maketitle

Weakly interacting massive particles~(WIMPs) are favored as particle candidates for non-baryonic cold dark matter~(DM), as their interaction strength can explain the thermal relic abundance of DM~\cite{Steigman:1984ac,Jungman:1995df,Bertone:2004pz}.
A promising way to identify a WIMP DM particle is to search for excess neutrino flux generated by WIMP self-annihilations inside the Sun~\cite{Press:1985ug,Silk:1985ax,Gaisser:1986ha,Gould:1987ir} (``WIMP neutrinos''). As the Sun travels on the Milky Way arm, WIMPs in the DM halo could occasionally become gravitationally bound after losing energy by scattering off nuclei in the Sun. The WIMPs then pair-annihilate in the deep solar core, producing neutrinos from decays of the annihilation products which propagate outward through the Sun and may be detected in terrestrial neutrino detectors.
An excess of a high-energy neutrinos from the Sun, with energies much greater than the $\sim$$O$(10 MeV) solar-fusion neutrinos, observed in neutrino detectors can be interpreted in terms of WIMP annihilation rate.
We assume that WIMPs annihilate via a single channel to a pair of fermions or bosons. Also, we assume that the WIMP capture and annihilation rates are in equilibrium which enables us to remove the WIMP-annihilation cross-section dependence and allows comparison to direct detection results on WIMP-nucleon elastic scattering cross section.

Thanks to the hydrogen-rich composition and large gravity of the Sun, tight limits on the spin-dependent~(SD) scattering cross section of WIMPs on protons have been placed by neutrino telescopes such as Super-Kamiokande~\cite{Tanaka:2011uf}, IceCube~\cite{Aartsen:2012kia}, Baksan~\cite{Boliev:2013ai} and ANTARES~\cite{Adrian-Martinez:2013ayv}. 
Recent event excesses or annual modulation signals reported by direct detection experiments such as DAMA/LIBRA~\cite{Bernabei:2008yi}, CoGeNT~\cite{Aalseth:2010vx}, CRESST~\cite{Angloher:2011uu}, CDMS~II Si~\cite{Agnese:2013rvf} and conflicting null results from other direct- and indirect-detection experiments motivate a careful search for light WIMPs below 30 GeV/$c^2$.
As typical resulting neutrino energies are roughly one third to one half of the WIMP mass~\cite{Jungman:1995df} or lower, the Super-Kamiokande~(SK) detector's sensitivity to few-GeV neutrinos makes it suitable for this search~\cite{Hooper:2008cf,Feng:2008qn,Kappl:2011kz,Rott:2011fh}.

\begin{figure*}[t]
\centering
\includegraphics[width=0.9\textwidth]{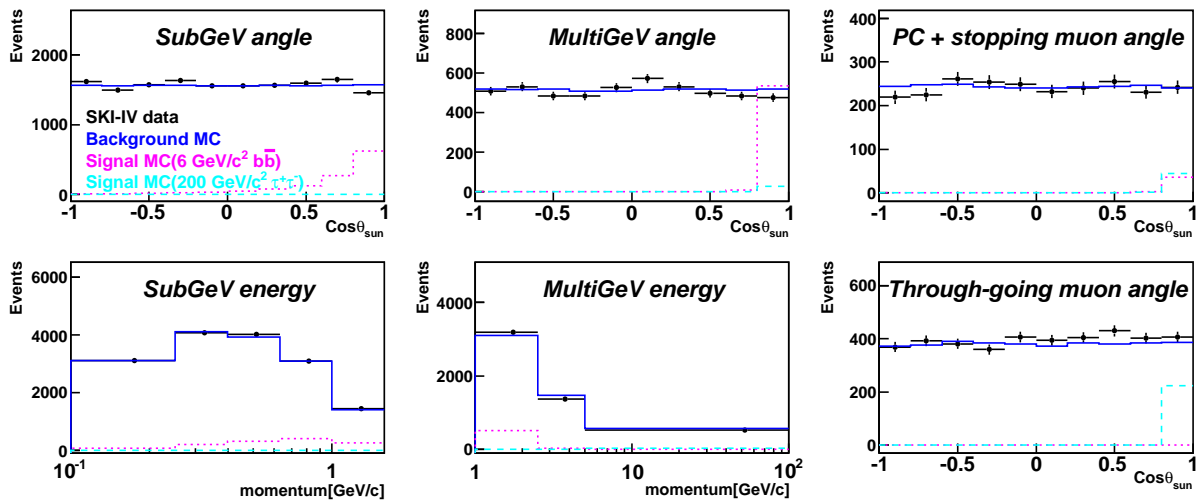}
\caption{Angular and reconstructed momentum [GeV/c] distributions of SK I-IV data (black crosses); atmospheric-neutrino background MC (normalized to data live time, blue solid); WIMP neutrino signal MC for the 6-GeV/$c^2$ $b\overline{b}$ channel (magenta dotted) / 200-GeV/$c^2$ $\tau^+\tau^-$ channel (cyan dashed) at 90\% upper limit, magnified 30 times for visibility.
Among seven sub-GeV samples, the single-ring e-like 0-decay-electron and $\mu$-like 0,1-decay-electron samples are combined and shown in the left column. In the middle column,  six multi-GeV samples are combined and shown. PC and up-$\mu$ samples are shown in the right column.
}
\label{fig2}
\end{figure*}

Super-Kamiokande is a cylindrical water Cherenkov detector located in the Kamioka mine in Japan. The inner detector (ID) of 22.5 kton fiducial volume is instrumented with 11,129 20-inch Hamamatsu photomultiplier tubes (PMTs) and the outer detector (OD) is instrumented with 1,885 8-inch PMTs for use as a veto.
Information about the experimental setup of the detector and its calibration, data reduction and event reconstruction can be found elsewhere~\cite{Fukuda:2002uc,Ashie:2005ik,Abe:2013gga}.

The SK high-energy neutrino data include neutrino events with visible energy$\textgreater$30~MeV and are divided into three categories: among contained events in which the neutrino interacts inside the ID,
fully-contained (FC) events have observed Cherenkov light entirely contained in the ID, partially-contained (PC) events additionally have an exiting particle that deposits energy in the OD. Upward-going muons (up-$\mu$) are produced by neutrino interactions in the rocks and water surrounding the detector.
FC events are distributed in seven sub-GeV (visible energy $<$ 1.33 GeV) and six multi-GeV (visible energy $>$ 1.33 GeV) sub-categories, and then further divided based on particle identification (e-like/$\mu$-like), number of reconstructed Cherenkov rings, number of decay electrons and so on.
PC events are classified as ``OD stopping'' or ``OD through-going'' based on their energy deposition in the OD.
Up-$\mu$'s either stop in the detector (stopping) or pass through the detector (through-going). Through-going up-$\mu$ events are further divided into ``showering'' and ``non-showering'' categories~\cite{Desai:2007ra}.

A signal from a 10-GeV/$c^2$ WIMP will mostly fall in the FC sub-GeV and multi-GeV samples, and heavier WIMP signals will be mainly distributed in the PC and up-$\mu$ categories.
To increase signal acceptance for light WIMPs, FC and PC events are included in the WIMP search sample. We take into account all neutrino flavors ($\nu_{\mu}$, $\overline{\nu}_{\mu}$, $\nu_e$, $\overline{\nu}_e$, $\nu_{\tau}$ and $\bar{\nu}_{\tau}$).
Compared to the previous analysis~\cite{Tanaka:2011uf} which used up-$\mu$ events only, the signal acceptance has increased 47 times for the 10-GeV/$c^2$ $b\overline{b}$ channel.

The analysis in this paper uses data accumulated during the SK I-IV run periods. The SK I (1996-2001), II (2002-2005, with half PMT coverage), III (2006-2008) periods correspond to 1489, 799, 518 days of live time for the FC/PC samples, and 1646, 828, and 636 days of live time for the  up-$\mu$ sample, respectively. The SK-IV period started in 2008; this analysis uses 1096.7 live-days of FC/PC/up-$\mu$ data collected until March 2012.

The Monte Carlo simulation~\cite{Hayato:2002sd,Ashie:2005ik} originally generated for the primary atmospheric neutrino flux~\cite{Honda:2011nf}
is divided into two independent samples and used for two purposes: first, to predict atmospheric-neutrino background, and second, to produce the WIMP neutrino signal by reweighting to match the signal flux and spectrum.
To predict the atmospheric-neutrino background, a 250-year MC sample per SK run period is normalized to the live time of each SK run period and oscillated with parameters: $\sin^2 \theta_{13} = 0.025$, $\sin^2 \theta_{12} = 0.304$, $\sin^2\theta_{23} = 0.425$, $\Delta m^2_{21} = 7.66\times 10^{-5}$ eV$^2$ and $\Delta m^2_{32} = 2.66 \times 10^{-3}$ eV$^2$.
In the other 250 years of MC sample per run period, events coming approximately from the solar direction are selected and then weighted by the ratio of WIMP neutrino flux to original atmospheric neutrino flux.
To simulate the neutrino flux from WIMP annihilation in the Sun at SK, we used the WIMP MC simulator WimpSim 3.01~\cite{Blennow:2007tw}.
Focusing on the light WIMP search, we considered WIMP masses from 4~GeV/$c^2$ to 200~GeV/$c^2$ for $\chi\chi\rightarrow\tau^+\tau^- / b\overline{b} / W^+W^-$ channels. The spectrum of the $b\overline{b}$ channel is ``softest" and the $\tau^+\tau^-$ channel produces a ``harder" spectrum.

In order to discriminate the large background of atmospheric neutrinos at low energies, which have an $\propto E^{-2.7}$ power-law spectrum, and to take into account the energy-dependent angular correlation for low-energy events, we perform a least-squares fit which makes full use of the energy, angle and flavor information.
Under the hypothesis that the collected SK data consist of both atmospheric neutrinos and WIMP neutrinos, we compare data to MC.
The data and MC in the 18 sub-categories are distributed into 480 bins based on reconstructed momentum and $\cos\theta_{sun}$, where $\theta_{sun}$ is the angle between the Sun and the reconstructed event direction.
Figure~\ref{fig2} shows an example of data distributions compared to the expected WIMP-induced signal.
We fit the data using the least-squares technique based on Poisson statistics to find the amount of signal contribution added to background which best matches the data. For each mass and annihilation channel, the energy spectrum and flavor composition of signal neutrino flux are fixed and the global normalization of the signal flux is allowed to vary freely.
The ``pull'' approach allows us to incorporate systematic uncertainties~\cite{Fogli:2002pt} in the fitting as:

\begin{center}
\begin{widetext}
$\chi^2 = \min\limits_{\{\epsilon_j\}}\left[ 2\sum\limits_{i=1}^{N} \{ N_i^{exp} - N_i^{data} + N_i^{data} ln(N_i^{data}/N_i^{exp}) \} + \sum\limits_{j=1}^J{( \frac{\epsilon_j}{\sigma_j})}^2 \right]$,\\
where $N_i^{exp}=N_{i}^{BG}(1+\sum\limits_{j=1}^J f_j^i \epsilon_j)+\beta N_i^{\chi}(1+\sum\limits_{j=1}^J g_j^i \epsilon_j),$
\end{widetext}\label{eq:chi2}\end{center}

and $i$ is the index of the bins; $N_i^{data}$ is the number of events observed in each bin; $N_i^{BG}$ is the background expectation in the bin; $N_i^{\chi}$ is the number of events of signal MC in the bin, where $\beta$, the global normalization parameter for signal, stands for the allowed fraction of signal;
$\sigma_j$ is the 1$\sigma$ value of the $j$-th systematic uncertainty; $\epsilon_j$ is its ``pull'' and $f^i_j$/$\sigma_j$ ($g^i_j$/$\sigma_j$) is the predicted fractional change of the number of background (signal) events in the $i$-th bin due to a 1$\sigma$ change of the $j$-th systematic uncertainty.
In the pull approach, $\chi^2$ is estimated by solving $\delta\chi^2/\delta\epsilon_{j}=0$ for all $\epsilon_j$'s~\cite{Fogli:2002pt}.

For background (signal), 66 (48) sources of systematic uncertainties are considered.
A total of 16 systematic uncertainties related to neutrino interaction in SK are considered in common for both background and signal. Another 14 uncertainty sources related to event reduction, and 11 sources related to event reconstruction and selection are calculated for SK I, II, III and IV independently and considered for both signal and background.
For background MC, a total of 25 uncertainties related to prediction of atmospheric neutrino flux, including 1$\sigma$ uncertainties on oscillation parameters, are considered.
All the uncertainties in neutrino interaction, event reduction, reconstruction, selection and background atmospheric neutrino flux are similar to those described in~\cite{Abe:2014gda} except the ones related to zenith angle distributions. These uncertainties are listed elsewhere (for example, Tab.~VI,VII,VIII in ~\cite{Abe:2014gda}).
For signal MC, 7 uncertainties related to the oscillation of WIMP neutrinos during propagation through the Sun, vacuum, and the Earth, are considered.
The full systematic uncertainty has up to 10\% effect on the WIMP sensitivity, where the largest contributions come from neutrino interaction and atmospheric neutrino flux uncertainties.

\begin{figure}[htb]
\centering
\includegraphics[width=0.5\textwidth]{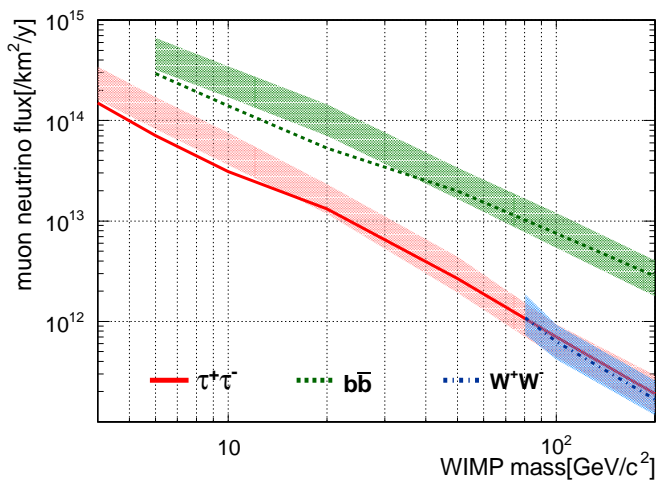}
\caption{The 90$\%$ upper limit on total integrated muon-neutrino flux from WIMP annihilations in the Sun at SK for the $\tau^+\tau^-$ channel in red solid; $b\overline{b}$ in green dashed; $W^+W^-$ in blue dot-dashed. The shadowed regions show 1$\sigma$ bands of the sensitivity study results (color scheme is the same as for data).}
\label{fig3}
\end{figure}

\begin{table*}[htb]
\begin{center}
\begin{tabular}{|p{1.3cm} p{1.8cm} p{1.3cm} p{1.2cm} ccccc|}
\hline
\hline
$m_{\chi}$ & annihilation & $\chi^2_{min}$ & $\Delta\chi^2$ & $\Delta\chi^2_{90}$ & $\nu_{\mu}$ ($\times 10^{12}$  & $\sigma_{SD,p}$ & $\sigma_{SI,p}$~($f_n$/$f_p$=1) & ($f_n$/$f_p$=$-$0.7) \\
(GeV/$c^2$) & channel & & at $\beta=0$ & (Bayesian) & km$^{-2}$y$^{-1}$) & ($\times 10^{-40}$cm$^2$) & ($\times 10^{-43}$cm$^2$) & ($\times 10^{-41}$cm$^2$) \\
\hline
4 & $\tau^+\tau^-$ & 508.1 & 0.87 & 4.4 & 150 & 2.22     & 87.3 &  24.5\\
\hline
6 & $b \overline{b}$ & 507.8 & 1.24 & 4.9 & 294    &  17.2        & 456 & 128\\
\cline{2-9}
 & $\tau^+\tau^-$ &  507.8 & 1.25 & 4.9 & 70.8 &         1.63  &44.5 & 12.2 \\
\hline
10 & $b \overline{b}$ & 507.4 & 1.65 & 5.4 & 140  &       14.9 &  240 & 67.5 \\
\cline{2-9}
& $\tau^+\tau^-$ & 507.6 & 1.40 & 5.1 & 31.0 &    1.31 &21.2 & 5.95\\
\hline
20 & $b \overline{b}$ & 507.4 & 1.65 & 5.4 & 53.1  &       14.3   & 120 & 44.8 \\
\cline{2-9}
& $\tau^+\tau^-$ &  509.0 & 0.06 & 2.4 & 13.2 &         1.42   &11.9 & 4.47 \\
\hline
50 & $b \overline{b}$ & 509.0 & 0.04 & 3.0 & 19.8  &       23.4   & 89.9 & 39.0 \\
\cline{2-9}
& $\tau^+\tau^-$ &  508.9 & 0.14 & 2.3 & 2.67 &         1.28   & 4.92 & 2.14 \\
\hline
80.3 & $W^+W^-$ &  508.8 & 0.17 & 2.3 & 1.09 &  3.13 & 8.26 & 3.73\\
\hline
100 & $b \overline{b}$ &  509.0 & 0.06 & 2.4 & 7.54 &  31.9 & 71.3 & 32.7\\
\cline{2-9}
& $W^+W^-$ & 508.9 & 0.14 & 2.3 & 0.63 &         2.80  & 6.26 & 2.87 \\
\cline{2-9}
& $\tau^+\tau^-$ &  508.9 & 0.16 & 2.3 & 0.70 & 1.24 & 2.76 & 1.26 \\
\hline
200 & $b \overline{b}$ & 508.9 & 0.12 & 2.3 & 2.81 &                44.9  & 63.4 & 30.4 \\
\cline{2-9}
& $W^+W^-$ &  508.9 & 0.07 & 2.4 & 0.17 &         3.00 &      4.23 &  2.03 \\
\cline{2-9}
& $\tau^+\tau^-$ & 508.9 & 0.08 & 2.4 & 0.19   & 1.33 & 1.88 & 0.90\\
\hline
\hline
\end{tabular}
\caption{$\chi^2_{min}$, $\Delta\chi^2$ at $\beta=0$, $\Delta\chi^2_{90}$ ($\Delta\chi^2$ for 90\% Bayesian upper limit), 90\% upper limit on the muon-neutrino flux from WIMP annihilations in the Sun at SK and SD/SI/SI~(IVDM) scattering cross section limits for each WIMP mass and annihilation channel.
}
\label{table1}
\end{center}
\end{table*}

The best-fit value of $\beta$, $\beta_{min}$, is defined as the value at which $\chi^2$ is minimized with respect to $\beta$.
For all tested WIMP hypotheses, we found the resulting $\beta_{min}$ values are all consistent with the hypothesis of no WIMP neutrino contribution, and we set 90\% upper limits assuming that the obtained $\chi^2$ values approximately follow a normal distribution.
To set a physically-meaningful confidence limit, we follow a Bayesian approach~\cite{Barnett:1996hr} and renormalise the $\beta$ distribution in the physically-allowed region ($\beta > 0$).
The fitted $\chi^2$ values at minimum ($\chi^2_{min}$; DOF is 479), the $\Delta\chi^2$ value at $\beta=0$ from the minimum, and the $\Delta\chi^2$ values for 90\% upper limit ($\Delta\chi^2_{90}$) calculated with the Bayesian approach are shown in Tab.~\ref{table1} for the tested WIMP mass and annihilation channels.

Figure~\ref{fig3} shows the derived 90\% upper limit on the muon-neutrino flux from WIMP annihilations in the Sun at SK. The limit shown here for $\nu_{\mu}$ is an example; limits on other flavors or antineutrinos are determined by appropriate flavor scaling for the given WIMP annihilation channel assumption.
The sensitivity study was done by substituting SK data with 500 sets of toy MC generated for the hypothesis of no WIMP-annihilation contribution and the result is shown together in Fig.~\ref{fig3} as a band representing a 1$\sigma$ range of results.

\begin{figure}[htb]
\centering
\includegraphics[width=0.5\textwidth]{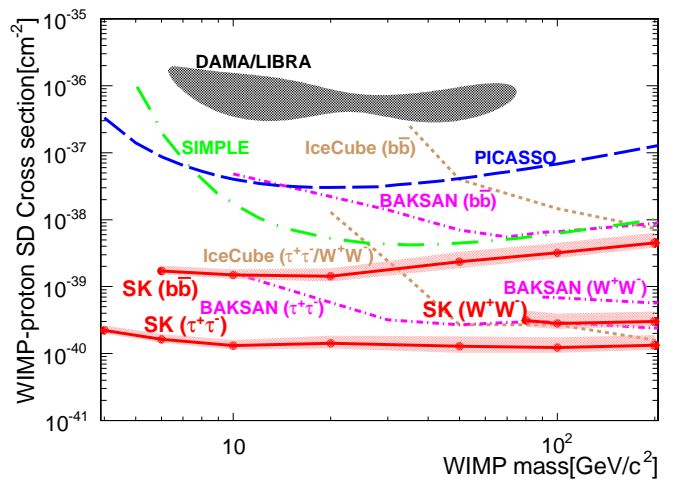}
\caption{90$\%$ C. L. upper limits on SD WIMP-proton cross section calculated at DarkSUSY~\cite{Gondolo:2004sc} default are shown in red solid with uncertainty bands to take account uncertainties in the capture rate for the $b\overline{b}$, $W^+W^-$ and $\tau^+\tau^-$ channels from top to the bottom. Also shown are limits from other experiments: IceCube~\cite{Aartsen:2012kia} in brown dashed: $b\overline{b}$ (top) / $W^+W^-$ or $\tau^+\tau^-$ (bottom); BAKSAN~\cite{Boliev:2013ai} in pink dot-dashed: $b\overline{b}$ (top) / $W^+W^-$ (middle) / $\tau^+\tau^-$ (bottom); PICASSO~\cite{Archambault:2012pm} (blue long-dashed); SIMPLE~\cite{Felizardo:2011uw} (green long dot-dashed). The black shaded region is the $3\sigma$ C.L. signal claimed by DAMA/LIBRA~\cite{Bernabei:2008yi,Savage:2008er}.}
\label{fig4}
\end{figure}

\begin{figure}
\centering
\includegraphics[width=0.5\textwidth]{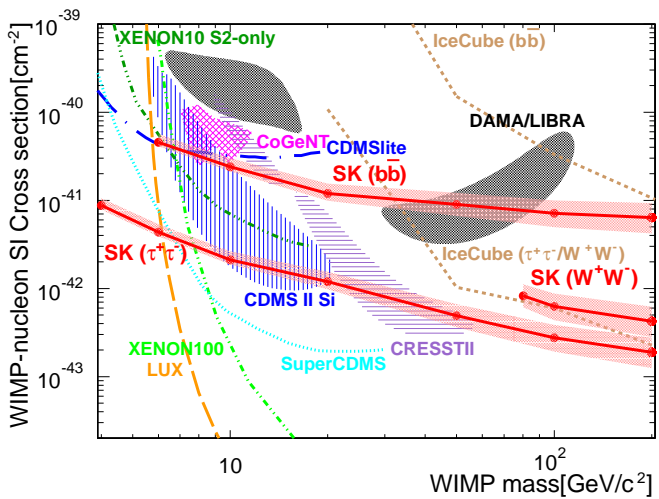}
\caption{90$\%$ C. L. upper limits on the SI WIMP-nucleon cross section (plotting scheme is the same as Fig.\ref{fig4}). Also shown are event excesses or annual modulation signals reported by other experiments: DAMA/LIBRA (black shaded regions, $3\sigma$ C.L.); CoGeNT~\cite{Aalseth:2010vx} (magenta diagonally cross-hatched region, 90\% C.L.); CRESSTII~\cite{Angloher:2011uu} (violet horizontally-shaded regions, $2\sigma$ C.L.); CDMS~II Si~\cite{Agnese:2013rvf} (blue vertically-shaded region, 90\% C.L.); and limits: IceCube~\cite{Aartsen:2012kia} in brown dashed: $b\overline{b}$ (top) / $W^+W^-$ or $\tau^+\tau^-$ (bottom); SuperCDMS \cite{Agnese:2014aze} (cyan dotted); CDMSlite \cite{Agnese:2013jaa} (blue long dot-dashed); XENON10 S2-only~\cite{Angle:2011th} (dark green dash triple dot); XENON100~\cite{Aprile:2012nq} (green dash double dot); LUX~\cite{Akerib:2013tjd} (orange long-dashed).}
\label{fig5}
\end{figure}

Using DarkSUSY 5.0.6~\cite{Gondolo:2004sc}, we convert the upper limit on the neutrino flux to upper limits on WIMP-nucleon cross sections.
We assume that WIMPs have only a single type of interaction with a nucleus, either an axial vector interaction in which WIMPs couple to the nuclear spin (SD) or a scalar interaction in which WIMPs couple to the nucleus mass (spin-independent, SI).
The SI coupling can have different couplings to neutrons ($f_n$) and protons ($f_p$). We consider two representative examples: the commonly-assumed isospin-invariant case ($f_n/f_p$ = 1) and the case of isospin-violating dark matter~(IVDM)~\cite{Chang:2010yk} with destructive interference $f_n/f_p =-$0.7~\cite{Feng:2011vu,Chen:2011vda,Frandsen:2013cna,DelNobile:2013cta}. The latter is calculated based on \cite{Gao:2011bq}.
Also assumed are a standard DM halo with local density 0.3 GeV/cm$^3$~\cite{Kamionkowski:1997xg,Yao:2006px}, a Maxwellian velocity distribution with an RMS velocity of 270 km/s and a solar rotation speed of 220 km/s.
The results are listed in Tab.~\ref{table1}. They are also plotted together with other experimental results in Fig.~\ref{fig4} for SD coupling and Fig.~\ref{fig5} for SI coupling for the isospin-invariant case.  For IVDM, we note that for the result given in the Tab.~\ref{table1}, the entire CDMS~II Si~\cite{Agnese:2013rvf} 90\% C.L. signal region will be in tension with our  $\tau^+\tau^-$ channel result.

There are several sources of uncertainties related to the WIMP capture process. Uncertainty in the composition of the Sun is considered by comparing the DarkSUSY default choice BS2005-OP model with the BS2005-AGS,OP model~\cite{Bahcall:2004pz} with lower heavy element abundances. We compared the Helm-Gould nuclear form factor~\cite{Gould:1987ir} used in DarkSUSY to the choices in \cite{Wikstrom:2009kw} and \cite{Ellis:2009ka}. These effects are predicted to be small in the SD-coupling case and up to 25/45\% in the SI-coupling where heavier elements than hydrogen contribute to the capture.
The effect of uncertainties in the velocity distribution of WIMPs is determined in \cite{Choi:2013eda} to be up to 40(25)\% for SD(SI) couplings.
The effects of the planets on the capture rate is determined to be negligible~\cite{Gould:1990ad,Sivertsson:2012qj}. Solar evaporation is expected to have no impact above a WIMP mass of 4~GeV/$c^2$~\cite{Griest:1986yu, Gould:1987ju, Hooper:2008cf, Busoni:2013kaa}.
These uncertainties are added in quadrature and indicated by the shadowed regions in Fig.~\ref{fig4} and Fig.~\ref{fig5}.
Uncertainty in the local WIMP density will make a similar vertical shift of the limits for all direct and indirect detection experiments, and so is not indicated.

In conclusion, the result of the first WIMP search using contained events in SK is presented. No significant signal excess was found for 4--200-GeV/$c^2$ WIMP hypotheses. The derived upper limit on the SD WIMP-proton cross section places the most stringent constraint to date for WIMP masses below 200 GeV/$c^2$ even for the softest ($b \overline{b}$) channel, assuming the equilibrium condition between capture and annihilation rates.
For the SI-coupling case with 100\% annihilation to $\tau^+\tau^-$, we exclude new regions for WIMP masses below 6 GeV/$c^2$.

\begin{acknowledgments}
We gratefully acknowledge cooperation of the Kamioka Mining and Smelting Company. The Super-Kamiokande experiment was built and has been operated with funding from the Japanese Ministry of Education, Culture, Sports, Science and Technology, the U.S. Department of Energy, and the U.S. National Science Foundation. 
\end{acknowledgments}
\bibliography{refs}
\end{document}